\documentclass[conference]{IEEEtran}
\usepackage[letterpaper, margin=.97in,top= 0.725 in,right=0.64in, left=0.64in]{geometry}

\usepackage[T1]{fontenc}
\usepackage[utf8]{inputenc}

\usepackage{graphicx,amsmath}
\ifCLASSOPTIONcompsoc
    \usepackage[caption=false, font=normalsize, labelfont=sf, textfont=sf]{subfig}
\else
\usepackage[caption=false, font=footnotesize]{subfig}
\fi

\usepackage[noadjust]{cite}
\usepackage[algo2e,ruled]{algorithm2e}
\newcommand{\ie}{\textit{i}.\textit{e}. }
\usepackage{setspace}

\usepackage{amsthm}

\theoremstyle{definition}
\newtheorem{definition}{Definition}[section]

\theoremstyle{remark}

\begin{document}

\title{Towards AoI-aware Smart IoT Systems}


\author{
	{
		Hasan Burhan Beytur,
		Sajjad Baghaee,
		Elif Uysal
	}	\\
	
	{Department of Electrical and Electronics Engineering, Middle East Technical University, Ankara, Turkey}\\

	hasan.beytur@metu.edu.tr,
	sajjad@baghaee.com,
	uelif@metu.edu.tr
}

\maketitle

\begin{abstract}
Age of Information (AoI) has gained importance as a Key Performance Indicator (KPI) for characterizing the freshness of information in information-update systems and time-critical applications. Recent theoretical research on the topic has generated significant understanding of how various algorithms perform in terms of this metric on various system models and networking scenarios. In this paper, by the help of the theoretical results, we analyzed the AoI behavior on real-life networks, using our two test-beds, addressing IoT networks and regular computers. Excessive number of AoI measurements are provided for variations of transport protocols such as TCP, UDP and web-socket, on wired and wireless links. Practical issues such as synchronization and selection of hardware along with transport protocol, and their effects on AoI are discussed. The results provide insight toward application and transport layer mechanisms for optimizing AoI in real-life networks.
\end{abstract}

\IEEEpeerreviewmaketitle

\section{introduction}
More and more, new IoT applications are taking place in our daily life. It is not surprising anymore that everyday objects as household appliances, vehicles, lights, waste containers, etc., are connected to the Internet. Different from usual data traffic, the IoT devices are usually generating small packets carrying status updates. These status updates could be samples of a slowly changing process such as the temperature or soil humidity, or a rapidly changing process such as a vehicle’s instantaneous acceleration or position, or the state of a networked control system \cite{SajjadBAGHAEE2015}. Therefore, the excessive number of IoT devices have vastly varying requirements for the network infrastructure.






Classical metrics such as throughput and latency are not appropriate for measuring quality of service for IoT and other status-update based applications or designing new service policies for them. In contrast to classical data communication, for status updating, the packets are not equally important - an old (out of order) packet should be discarded, if the newest update is critical. Often, the IoT applications require \emph{sufficiently timely information} to use in computation or actuation at the end node. Providing sufficiently timely data to a growing number of IoT devices is currently a challenge depending on communication protocols to be re-thought completely. However, to tackle this challenge the concepts of timeliness should be quantified. A useful metric that emerged in recent years is Age of Information \cite{6195689}, which refers to the time elapsed since the most recent update is received at the destination. 

Time average age, expected age, peak age, and the distribution of age in general (e.g., Age Violation Probability) have been analyzed for various service policies and optimized under various assumptions for the packet generation process (see ~\cite{ageViolation_Durisi} and references therein). It has been shown that in networked control and situational awareness, end-to-end performance is captured by a penalty function that is age itself or an increasing function of age~\cite{Klugel2019}.

In the literature, many works are investigating the average age analytically in different queuing systems \cite{6195689,Kaul_Yates_Gruteser_2012,Tanaka_2018}. In \cite{Kaul_Yates_Gruteser_2012}, the advantages of LCFS over FCFS is investigated and results are proposed for preemptive and non-preemptive LCFS system. In addition to these, \cite{Tanaka_2018} is a reference study summarizing all existing results and providing bounds and closed-form expressions for average age for the general case. But the theoretical results are hardly helping to understand the age behavior in real systems such as the Internet. Few works are discussing the age characteristics in realistic systems, one of them is \cite{8433695}, providing the AoI measurements on real testbed over TCP/IP links served by WiFi, LTE, 3G, 2G, and Ethernet. The provided results in \cite{8433695} have similar U-shaped characteristic comes across the analytical results of FCFS systems with Poisson or Gamma distributed arrivals \cite{gammaAwakening}. 

The ability to decide in large quantities of data and large-scale communication network without a priori knowledge about the network statistics is a particular benefit of machine learning approaches. There have been studies applying machine learning techniques for AoI-aware scheduling and sampling. The idea of waiting before sampling proposed in \cite{UpdateOrWait} is combined with reinforcement learning in \cite{EgemenRL,ClementRL} to perform smart sampling without having any prior knowledge about network. Similarly, in \cite{BeyturRL,Ceran_INFOCOM2019} age-aware scheduling is studied using RL methods. 


In this work, while interpreting the age measurements obtained on our testbeds using TCP and UDP protocols, we benefit from the theoretical results provided for FCFS systems. Additionally, general issues arising while using age metrics in realistic setups such as the age bias caused by synchronization errors, are explored. To the best of our knowledge, this paper presents the first reported investigation of AoI on real IoT testbeds. The ideas and results provided in \cite{siu2019, 8433695} are explored in more depth in this work.

The rest of this paper is organized as follows: Section \ref{aoi-intro} introduces Age of information (AoI). Effect of synchronization error on AoI is investigated in  Section \ref{sync-aoi}, and AoI measurements on real-life systems and their results are demonstrated in Section  \ref{AoI-measurements}. Finally, conclusions are discussed in section \ref{conclusion}.


\section{Age of Information: definition}
\label{aoi-intro}

Age of information is a new metric, which quantifies the timeliness of a status update process transmitted from a source to a destination. 

\begin{figure}
	\centering
	\includegraphics[width=0.9\linewidth]{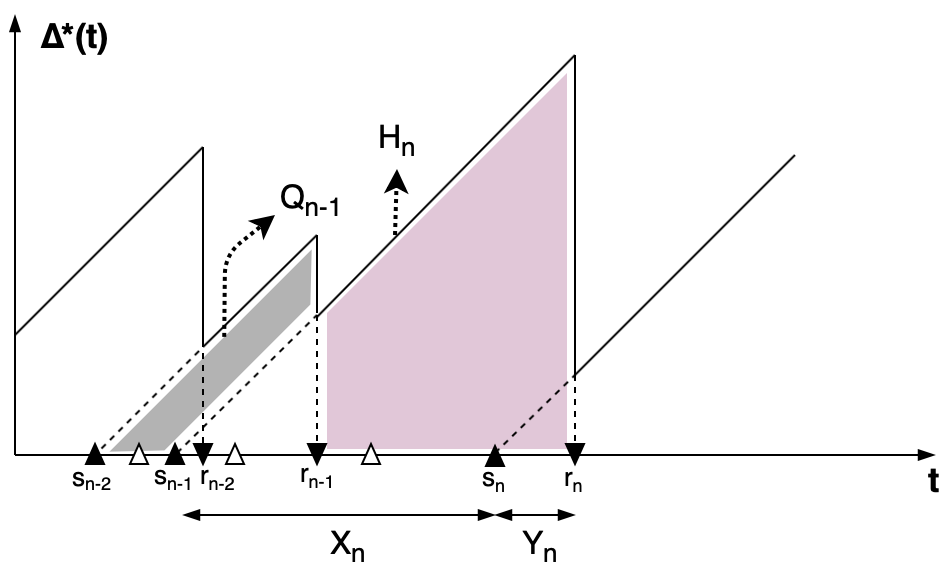}
	\caption{Sample path of the age process $\Delta(t)$}
	\label{aoi1}
\end{figure}

The status age $\Delta(t)$ is defined as the time that has elapsed since the newest data packet available at the destination at time $t$ was generated at the source. More precisely, $\Delta(t) = t - U(t)$,  where $U(t)$ is the generation time (i.e. time stamp) of the newest data that the destination has received by \mbox{time $t$}. This definition leads age to follow a sawtooth pattern as in the sample path given in Fig. \ref{aoi1}. The example in Fig. \ref{aoi1} assumes a case where the observation begins at $t=0$ with an empty queue at the destination and $\Delta(0) = \Delta_{0}>0$.  The source generates status updates at $s_{1}$, $s_{2}$, $\cdots$, $s_{n}$, which are received at $r_{1}$, $r_{2}$, $\cdots$, $r_{n}$, respectively. In the absence of any updates, the status age increases linearly in time and decreases just after an update is received. The area under the age graph normalized by \mbox{time $T$} gives the time average of AoI.

\begin{equation}
\overline{\Delta} = \frac{1}{T}\int_{0}^{T} \Delta(t) dt
\label{eq:aoi}
\end{equation}

For $n$ transmitted packets, the area is composed of the area of polygon $Q_1$, isosceles trapezoids $Q_i$'s for $2\leq i \leq n$ and the triangle of length $Y_n$ positioned at the bottom of $Q_n$.

\begin{equation}
\overline{\Delta} = \frac{Q_1 + \sum_{i=2}^{N(T)} Q_i + Y_n^2/2}{T}
\label{eq:age with q}
\end{equation}


\begin{equation}
    \lim_{T\to\infty} \overline{\Delta} = \frac{1}{T} \sum_{i=2}^{N(T)}Q_i,
\end{equation}


where, $Q_i = \frac{1}{2}(2r_i - s_i - s_{i-1})(s_i - s_{i-1})$ and $N(T)=max\{n|r_{n} \leq T\}$, the maximum number of arrival updates by time $T$. Note that, $(s_i - s_{i-1})$ is $X_i$, i.e. the inter-arrival time between successfully transmitted packets.

The time average AoI can also be expressed in terms of $H_i$'s, as:

\vspace{-0.05in}
\begin{equation}
\overline{\Delta} = \frac{1}{T} \sum_{i=1}^{N(T)} H_i
\label{eq:age with h}
\end{equation}

\vspace{-0.05in}
\begin{equation}
H_i = (r_i - r_{i-1})(r_{i-1}-s_{i-1}) + \frac{(r_i - r_{i-1})^2}{2}
\label{eq:hs}
\end{equation}

Note that, $(r_i - r_{i-1})$ is the inter-departure time between $(i-1)^{th}$ and $i^{th}$ packets  and $(r_{i-1}-s_{i-1})$ equals to $Y_{i-1}$ which is the system time of $(i-1)^{th}$ packet.

The AoI-aware scheduling and control algorithms require the calculation of AoI on receiver or transmitter, depending on the capabilities of the devices or according to where control decisions will be made. For example, if it is required to measure average AoI of a connection between a simple sensor and a server, AoI can be calculated on the receiver. On the other hand, if the transmitter is adaptively changing the sampling rate to achieve the minimum average AoI possible, then the transmitter needs to know the AoI, so in this case the AoI is measured at the transmitter side. 

Another metric investigated in the AoI literature is the time average of peak age of information.
\vspace{-0.05in}
\begin{equation}
    \overline{\Delta_{peak}} = \frac{1}{N(T)} \sum_{i=1}^{N(T)}\Delta(r_i) 
\end{equation}

The time average can be calculated using arrival and departure instances of packets
\vspace{-0.05in}
\begin{equation}
    \overline{\Delta_{peak}} = \frac{1}{N(T)} \sum_{i=1}^{N(T)} (r_i - s_{i-1})
    \label{eq:paoi}
\end{equation}

In the following section, we analyze the effect of synchronization error on average age for general age penalty functions \cite{UpdateOrWait,NonlinearAge_Kosta}. For this analysis, a more general formulation of average age is made using integral of $f(\Delta(t))$.


\vspace{-0.05in}
\begin{equation}
    \frac{1}{T} \int_{0}^T f(\Delta(t)) dt = \frac{1}{T} \sum_{i=1}^{N(T)} \int_{r_{i-1}}^{r_i} f(\Delta(t)) dt
    \label{eq:aoipenalty}
\end{equation}

This notation can be simplified using Definition \ref{corollary:penalty}.

\begin{definition} For a measurable, non-negative and non-decreasing age penalty function $f(\Delta(t)) : [0,\infty] \to [0,\infty]$, the time average age penalty is 
\vspace{-0.05in}
\begin{equation}
    \frac{1}{T} \int_{0}^T f(\Delta(t)) dt = \frac{1}{T} \sum_{i=1}^{N(T)} F(r_i - s_{i-1}) - F(r_{i-1}-s_{i-1})
    \label{eq:aoipenalty_corollary}
\end{equation}
where $F(t) = \int_0^t f(\tau) d\tau$, and $s_i$'s and $r_i$'s are packet generation and receiving times, respectively.
\label{corollary:penalty}
\end{definition}

\section{Effect of synchronization error on AoI}
\label{sync-aoi}
As mentioned in the previous sections, for AoI computation we need to get time stamps from both receiver and transmitter. If the receiver and transmitter have their own system clocks, they will have synchronization issue. There are several synchronization methods for networked systems, like Network Time Protocol and methods using GPS as reference, but none of these are perfect, and the synchronization error induces an error in age measurements.
In this section, the effect of the synchronization error in age is investigated.

If we neglect the time shift in the TX and RX clocks during the observation period, It can be assumed that the difference between two clocks is only a constant bias. With this assumption, there is a constant difference B between the transmitter and receiver clocks.
\vspace{-0.05in}
\begin{equation}
    t_{RX} = t_{TX} + B
\label{eq:clockbias}
\end{equation}

In the rest, the time stamp sampled with respect to distant time reference is shown with an apostrophe.
\begin{equation}
    r_i' = r_i + B
\label{eq:r_i'}
\end{equation}

When Definition \ref{corollary:penalty} is used, the age bias occurred due to synchronization error appears to be

\begin{equation}
    \overline{\Delta_{Bias}} = f(\overline{\Delta'}) - f(\overline{\Delta})
\end{equation}
\begin{multline}
    \overline{\Delta_{Bias}} = \frac{1}{T} \sum_{i=1}^{N(T)} F(r_i+B-s_{i-1}) - F(r_{i-1}+B-s_{i-1}) \\
     - F(r_i-s_{i-1}) + F(r_{i-1}-s_{i-1})
     \label{eq:agebias}
\end{multline}

\begin{figure}
	\centering
	\includegraphics[width=0.9\linewidth]{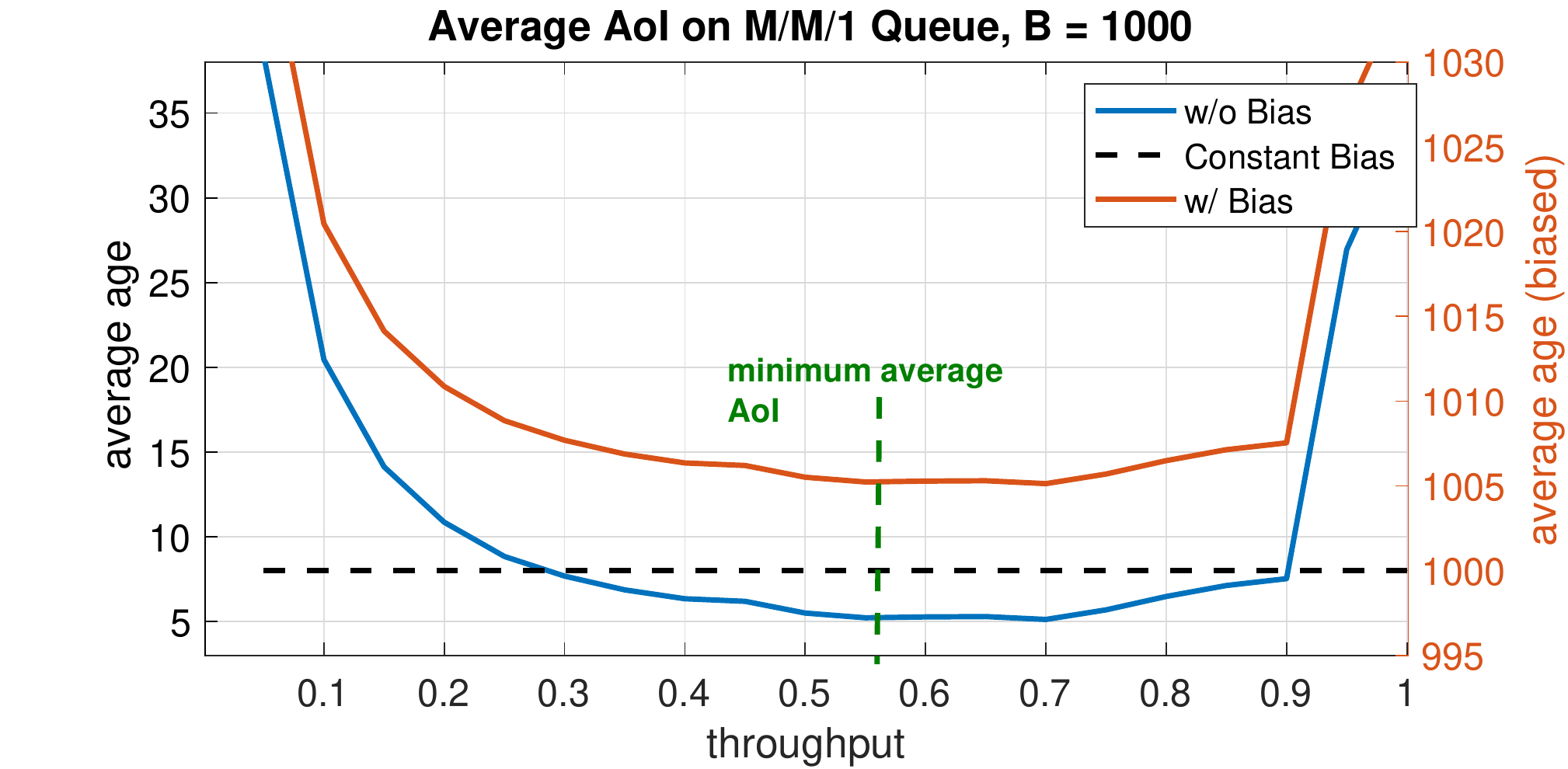}
	\caption{Average AoI measured using Monte Carlo simulation, an artificial synchronization error is added on the departure time stamps, $B=1000$, $\alpha = 1$}
	\label{constantBias}
\end{figure}


Next, we compute the age biases for the linear, exponential and logarithmic penalty functions, respectively. These penalty functions are commonly used in literature \cite{NonlinearAge_Kosta,Klugel2019}.
\begin{equation}
   f(t) =
\begin{cases} 
      \alpha t \\
      e^{\alpha t} -1 \\
      log(\alpha t +1) 
   \end{cases} 
   \label{functions}
\end{equation}


Using \ref{eq:agebias},



\begin{equation}
    \overline{\Delta_{Bias, Linear}} = \alpha B
    \label{eq:linear}
\end{equation}{}
\vspace{-0.05in}
\begin{multline}
     \overline{\Delta_{Bias, Exp}} = \frac{1}{\alpha \sum_{i=1}^{N(T)} (r_i - r_{i-1})} \Big( e^{\alpha (\theta + B)} - \\
     e^{\alpha (\beta + B)} - e^{\alpha \theta} + e^{\alpha \beta} \Big) 
\end{multline}

\begin{multline}
     \overline{\Delta_{Bias, Log}} = \frac{1}{\sum_{i=1}^{N(T)} (r_i - r_{i-1})} \bigg( \frac{1}{\alpha} \Big( log(\alpha \beta+1) - \\ log(\alpha \theta +1) +log(\alpha (B +\theta)+1) - log(\alpha (B +\beta)+1) \Big) - \\ \theta log(\alpha \theta +1) +\beta log(\alpha \beta +1) + log(\alpha (B + \theta)+1)(B + \theta) \\
     -log(\alpha (B + \beta)+1)(B + \beta)\bigg) 
\end{multline}
where $\beta = r_{i-1} - s_{i-1}$ and $\theta = r_i - s_{i-1}$.


In Figure \ref{constantBias}, the effect of clock difference between TX and RX on average age can be seen. For (\ref{eq:linear}) ($\alpha = 1$), where the synchronization error applies a constant shift in average age. Note that, when a non-linear age penalty function is used, the average age penalty measurements are distorted by the synchronization error, and the bias is not constant. This can lead to finding an undesirable operating point in terms of AoI. 

In the next section, the AoI measurements taken from real experimental setups are provided. Because of the issues about the nonlinear age penalty functions, we used the linear penalty functions with $\alpha = 1$.

 Before taking the measurements, the RTT between TX and RX is estimated sending several packets and receiving the ACKs. Because the ACK packet is very small, it is assumed that the transmission time of an ACK packet equals to the estimated RTT. Then, using this assumption we estimated the synchronization error. This method is not the best way to synchronize the transmitter and the receiver, but it is ensured that the age bias is bounded by the RTT.
 
 Note that, according to (\ref{eq:linear}), a constant bias exists due to the synchronization error. Hence it is unavoidable that our measurements contain an error, which is upper bounded by the value of the RTT. Hence, the values plotted represent absolute AoI up to a constant offset, the variation of the values within themselves being correct. 

\section{AoI Measurements on Real-life Systems}
\label{AoI-measurements}
In this section, the AoI measurement results taken from two experimental setups are discussed. The first setup investigates the age behavior of TCP and UDP flows over the Internet. The second setup examines a local, one-hop IoT connection. On a local IP network, we investigated the age behavior of TCP and UDP transmissions between IoT nodes. The second setup is more isolated, but mainly focuses on the effect of low-performance IoT devices on AoI.

In the upcoming sections, we will shortly discuss the TCP/UDP characteristics followed by focusing on the results obtained on our testbeds over the Internet and local connections.

\subsection{TCP/UDP Characteristics}
Both TCP and UDP are commonly used in many applications, and they are the most famous protocols. They have developed in many ways and have many versions and extensions, but in this work, only the general properties of them are in our focus.

TCP was developed to transfer large amounts of data, without loss. It was optimized to increase the throughput. To achieve this goal, it uses several mechanisms such as congestion control, adaptive window size, and re-transmission of lost packets. UDP is mainly used for the real-time applications which can tolerate occasional packet losses. UDP doesn't have any intelligent mechanisms like changing packet size or re-transmission. 

\begin{figure}
	\centering
	\includegraphics[width=0.9\linewidth]{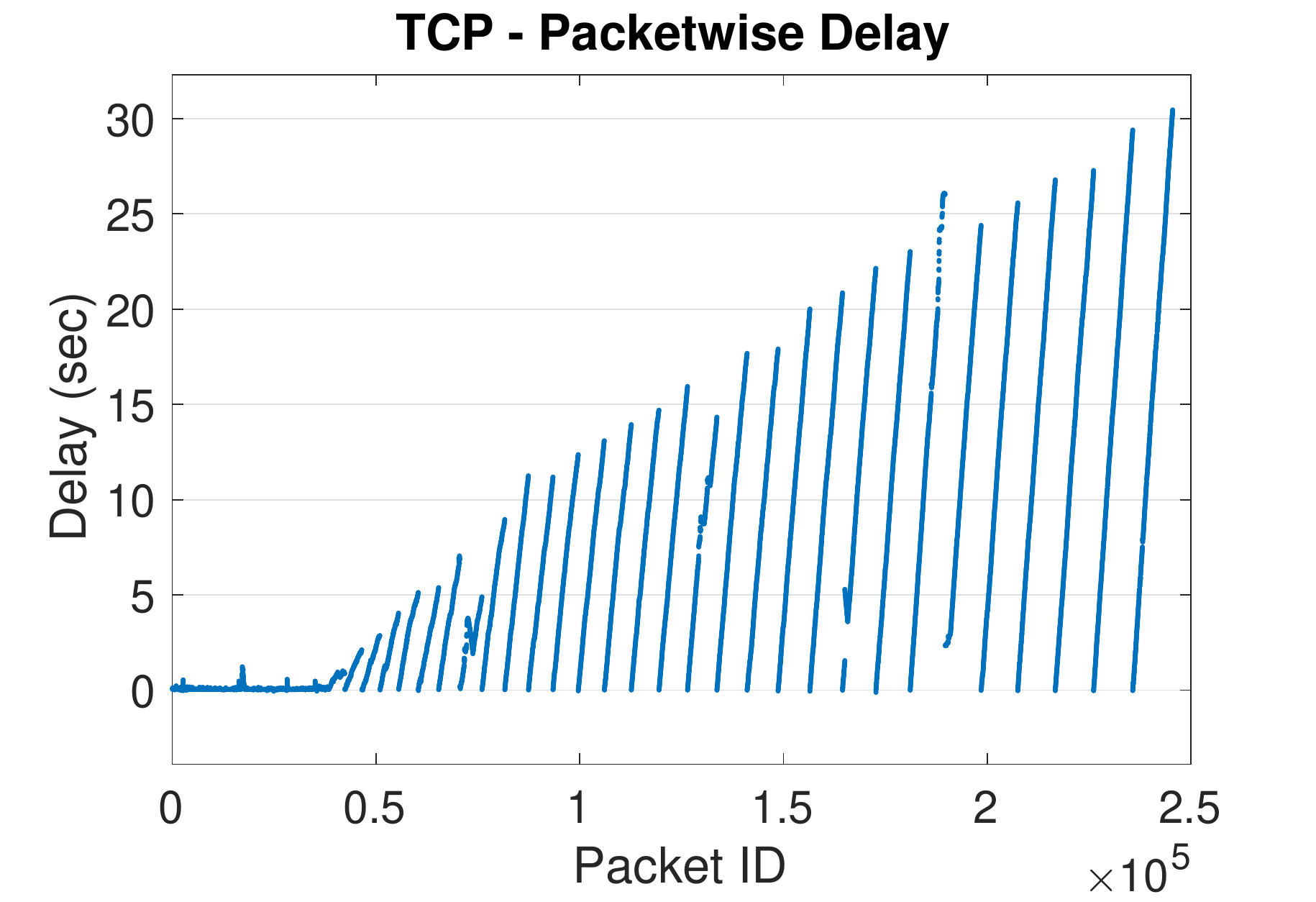}
	\caption{Packet-wise delay using TCP over Internet, connection established between Ankara and Istanbul/Europe with regular PCs}
	\label{tcpDelay}
\end{figure}

\begin{figure}
	\centering
	\includegraphics[width=0.9\linewidth]{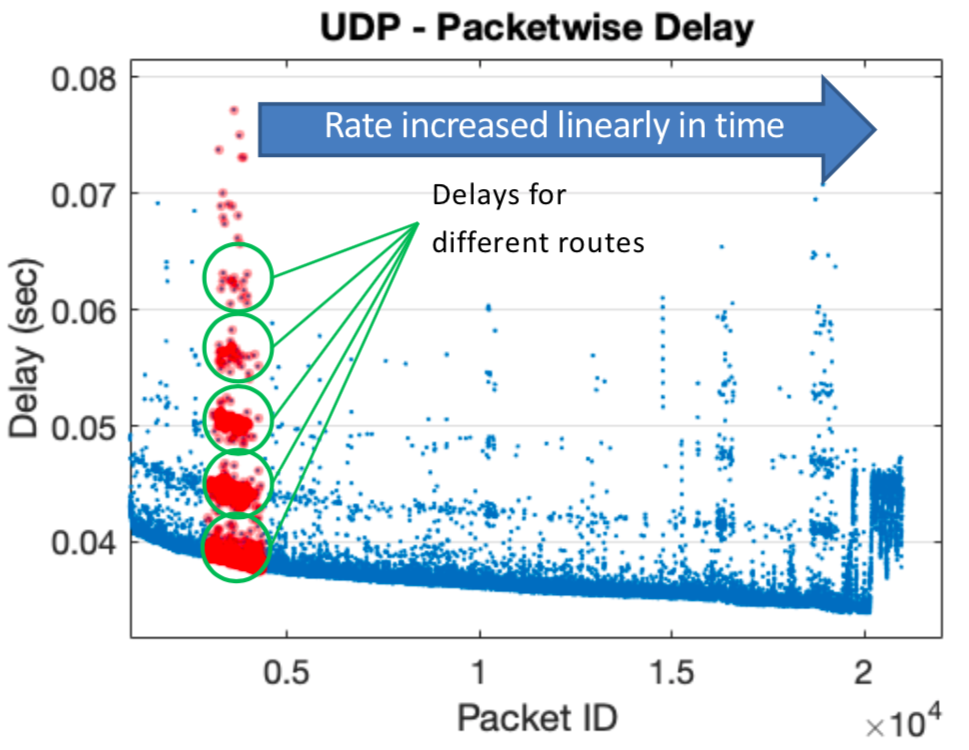}
	\caption{Packet-wise delay using UDP over Internet, connection established between Ankara and Istanbul/Europe with regular PCs}
	\label{udpDelay}
\end{figure}

Because of these main properties, there are several operational differences between TCP and UDP. Firstly, TCP is computationally more intense than UDP. In other words, due to the additional mechanisms, TCP requires larger number of CPU cycles per packet than UDP. This is also the reason why achievable throughput is higher when using UDP \cite{TCP-UDP}. Secondly, in the case of TCP, because of the re-transmission mechanism, the increasing congestion directly effects the delay of transmitted packets. As you can see in Figure \ref{tcpDelay}, increasing packet rate causes more congestion. While many packets are re-transmitted, the rest are waiting in the queue, both increase the packet delays. Since UDP does not have any re-transmission mechanism, delays of packets stay at a constant level. In Figure \ref{udpDelay}, several delay levels can be distinguished. We think that different levels belong to different transmission routes. You can also see that when the increasing rate exceeds the stable region, the delays jump to another level, but still stay constant, because packets will be getting discarded after this point.

\vspace{-0.in}
\subsection{Comparison between Computational Complexities of UDP and TCP}

As we mentioned earlier, unlike UDP, TCP has many control mechanisms to optimize throughput and compensate packet loss. The additional mechanisms of TCP increases computational overhead per packets. In the case of transferring huge amount of data via TCP, when all data is prompted to be sent at once, this computational overhead is dealt very well with kernel level operations. However, when new status updates are generated and sent over TCP at each time, the computational overhead becomes bottleneck limiting the maximum packet per second that can be processed. In our setup, with TCP,  the transmitter computer was able to reach to ~4000 packets per second. In case of UDP, the maximum packet generation rate was around 10000 packets per second. 

\subsection{AoI Behavior induced by the FCFS Service Policy}
As observed in \cite{6195689,Kaul_Yates_Gruteser_2012,Tanaka_2018}, in FCFS systems without any strict buffer management or limitation, at low throughput, the average age tends to decrease as throughput increases. After the queuing delay dominates the effect of frequent update rate, i.e. the communication system has difficulty to service the high throughput, the average age increases as throughput increases. Therefore, in all related studies, the U-shaped average age versus throughput figure exists. In today's network infrastructures FCFS buffers exist in routers, switches and access points. Therefore, the similar age characteristics on the Internet or IoT networks may be expected. 

Nonetheless, the experimental results of this study show that the properties of transport protocol and the TX/RX capabilities, i.e. CPU power, communication capabilities, etc., have significant effect on age behavior. The effects of selection of transport protocol are explained in details in following sections.

The effects related to the device's capabilities can be discussed in two steps. Firstly, the CPU should be able to generate enough number of packets per second to achieve low average age. Secondly, the communication module of the TX and RX should be able to process the generated packets. If one of them is not powerful enough, that one becomes the bottleneck for the age behavior. For example, if the CPU is not enough to push the limits of communication module or channel, the increase in average AoI at high rates is not observed. This factor is very important especially for low-power IoT devices. With the regular transport protocols, such as TCP and UDP, IoT devices have problems in terms of AoI. This shows us the necessity of a new age-aware transport protocol.

\begin{figure}[ht]
	\centering
	\includegraphics[width=1\linewidth]{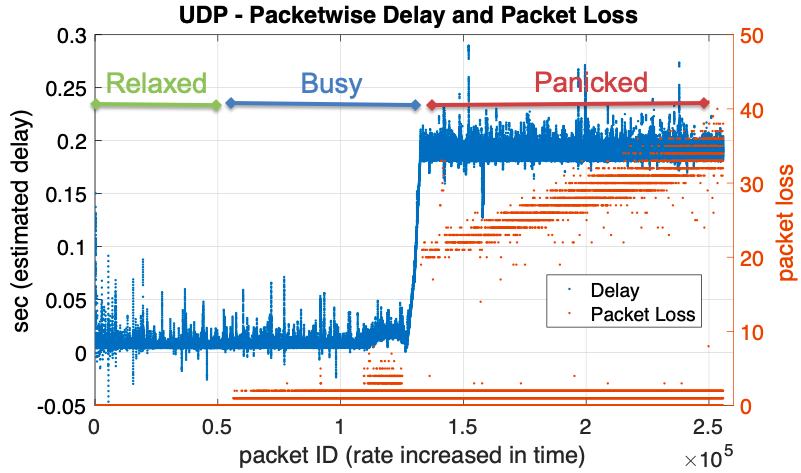}
	\caption{Packetwise delay with packet losses using UDP over Internet, using PCs}
	\label{udpInternet_delay}
\end{figure}

\begin{figure}[ht]
	\centering
	\includegraphics[width=0.9\linewidth]{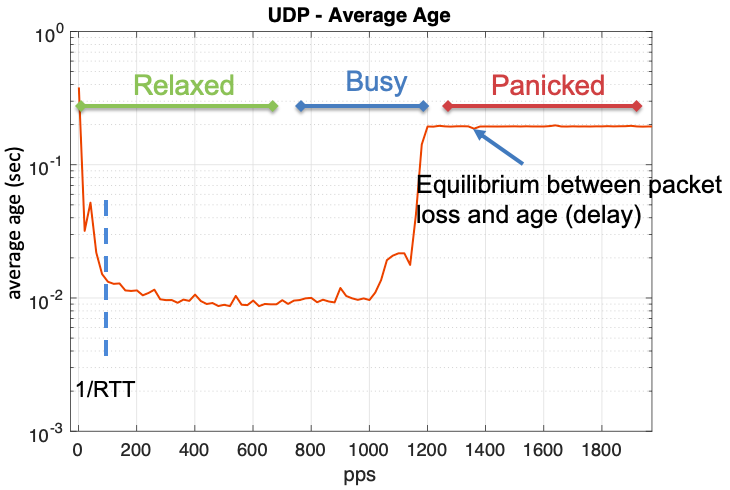}
	\caption{Average age using UDP over Internet, using PCs}
	\label{udpInternet_Age}
\end{figure}

\subsection{Experiment Setups}
 To compare the performance and understand the limitations, in this work, we have used two test setups entitled as Internet based Testbed with PCs and IoT based Testbed. In the Internet based Testbed, we established connection between high power desktop PCs to send TCP/UDP packets through regular Internet/IP infrastructure. Three PCs are located at Istanbul/Anatolia, Istanbul/Europe, and Ankara to test different paths with different path delays. The PC in Istanbul/Europe is the receiver node, and the other PCs send packets to it. The path between PCs in Istanbul/Anatolia and Istanbul/Europe has approximately 7 hops and 6 ms RTT The path between Ankara and Istanbul/Europe has approximately 12 hops and 80 ms RTT. The EchoServer has AMD Opteron 6174 (2.2 GHz) and the clients have Intel i5-8600K (3.6 GHz). 
In the second testbed, we built a local wireless IP network using a Wi-Fi router as the central node. In this setup, IoT devices (one is TX, other one is RX) send TCP/UDP packets each other through the central router node using 802.11n. The IoT devices we have used are NodeMCU ESP32 with Xtensa® LX6 (600 MIPS). Since this IoT device is not capable of running any operating system, the TCP/UDP operations are performed by Lightweight IP-Stack (LWIP), which is an open-source software and commonly used by different IoT devices.
With these test setups, we aimed to understand the effect of device capabilities on AoI and AoI behavior of most commonly used protocols TCP and UDP in both regular PCs and low-power IoT devices. 

\subsection{AoI on Internet}

In case of UDP, rather than delays incurred in transport layer queues, the increasing rate of packet loss at high load is the main factor increasing the average AoI. According to the experiment result, it seems that without loss of generality, the queuing delay is negligible.

In the Figure \ref{udpInternet_delay} and \ref{udpInternet_Age}, the results of UDP transmission tests using the Internet based testbed can be seen. In Figure \ref{udpInternet_delay}, the delays of successfully transmitted packets and packet losses occurring between the consecutive successful packets are shown. In this figure, although the horizontal axis indicates the Packet IDs, as the sample generation rate is  increased in time, it also illustrates the sample generation rate. In Figure \ref{udpInternet_Age}, the calculated average age values for the same experiments are shown. In this figure, the rate increase can be seen clearly. To make the connection between these figures, the operating regions are marked.

When we look at the UDP tests on the Internet based testbed, the operation can be divided into three regions, namely, Relaxed, Busy and Panicked. These regions are specified by the packet loss characteristics. In the Relaxed region, because of low transmission rate, the network is not congested. While operating in this region, increasing rate results in decreasing average age.
When the network begins to be congested the intermediate nodes along the transmission path randomly drop packets. As you can see in the Figure \ref{udpInternet_delay}, in this region the number of packet loss is generally less than 3 packets. In Busy region, the intermediate nodes still manage to compensate the high traffic. Note that, the packet-wise delays in Relaxed and Busy regions are at the same level. 
When the packet generation rate is increased further, the network is not able to tolerate high traffic anymore and eventually, high amount of packet losses occur. Interestingly, in the Panicked region, the packet-wise delays jump to a higher value and stay constant as rate increases. Consecutively, because of the equilibrium achieved between delay and packet loss, although the rate increased, the average age stays at a constant level. This occurs only in case of UDP, since UDP does not re-transmit the failed packets.

\begin{figure}[ht]
	\centering
	\includegraphics[width=0.9\linewidth]{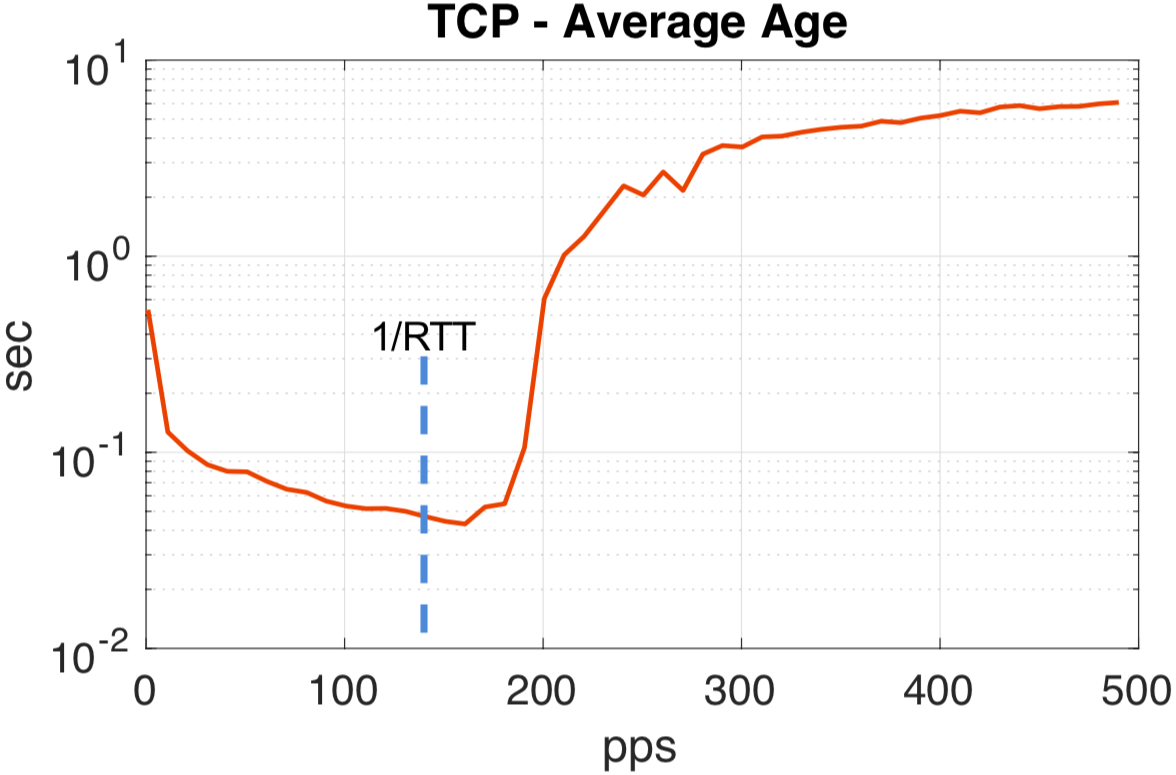}
	\caption{Average age using TCP over Internet, using PCs}
	\label{tcpInternet_Age}
	
\end{figure}

In the case of TCP, to ensure that all transmitted packets are successfully transmitted, the re-transmission mechanism is used. Therefore, in the experiments we did not observe any packet loss, as expected. However, the re-transmitting failed packets is costly in terms of AoI at high rates. In the Figure \ref{tcpInternet_Age}, the average age under increasing packet generation rate is illustrated. Note that, the y-axis is plotted in log-scale. Unlike UDP, in the case of TCP, due to re-transmissions mechanism, the increasing rate results in increase in delay. Consecutively, we get a plot similar to the U-shaped age-throughput graph of FCFS queues.

\vspace{-1pt}
\subsection{AoI on IoT}
The IoT modules are generally work at low power and they have very limited computational power. Under these circumstances, the achievable AoI characteristics are worth to be investigated. In our setup, a local Wi-Fi network is built using a router, and the ESP32 nodes have communicated with each other using TCP and UDP. \if The TCP/UDP compatibility is supplied to the IoT nodes with Lightweight IP-Stack (LWIP), which is developed by open-source community. \fi

\begin{figure}
	\centering
	\includegraphics[width=0.9\linewidth]{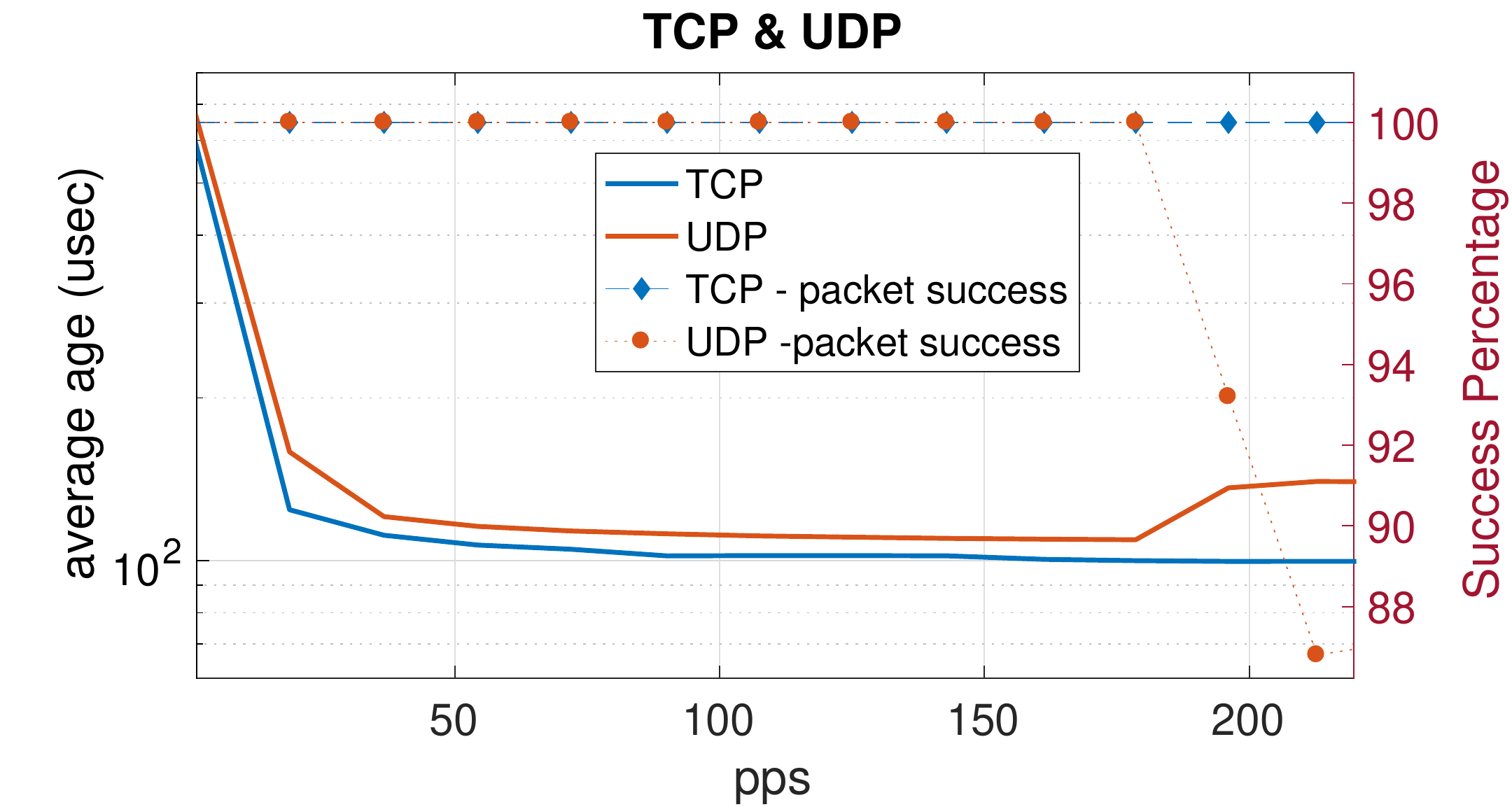}
	\caption{Average Age and Packet Loss using TCP and UDP over local Wi-Fi Network, using IoT devices}
	\label{udptcpIoT_age}
	\captionsetup{belowskip=0pt}
\end{figure}

According to our test result, AoI behavior on IoT devices is limited by device’s memory and computational power. The maximum packet generation rate is significantly lower than regular PCs and the available TX/RX buffer size is too small to observe the queueing delay occurring at high rates. As you can see in the Figure \ref{udptcpIoT_age}, due to the device’s limitations, the U-shaped AoI behavior is not observed clearly, for both UDP and TCP. Specific to our device, we observed that the Wi-Fi module integrated to ESP32 has higher throughput than the processing unit itself. Therefore, especially in the case of TCP, the device can generate less number of packets than it can transmit. With other IoT devices like ones using LoRa protocol to communicate, the AoI behavior can be observed better due to the limited bandwidth.

Another interesting result is 300 ms jitter observed while using UDP with LWIP, which is huge. In case of TCP tests, the delay was around  1-2 ms. We think that this wrong operation is caused by the wrong buffer management in UDP implementation in LWIP Stack. Therefore, until the bug is fixed, we suggest to use TCP rather than UDP on IoT devices.

\section{Conclusion}
\label{conclusion}
In this work, age behaviour in real-life connections including an IoT access link, and end-to-end UDP and TCP flows has been investigated. Practical computation of age and the bias arising due to synchronization error between transmitter and receiver have been discussed. Results from our experimental setups provide guidance for developing an AoI-aware transmission protocols.
\vspace{-0.05in}
\section*{Acknowledgments}
This work was supported by the Scientific and Technological Research Council of Turkey (TUBITAK) Project No. 117E215 and 117E003, and Huawei. The first author was funded by a Turk Telekom 5G Fellowship. We thank JeoIT, inc. for use of their equipment.

\bibliographystyle{IEEEtran} 
\bibliography{biblioList}

\end{document}